\documentclass{article}
\RequirePackage[]{amssymb}

\def\({\left(}
\def\){\right)}

\newcommand{\beq}{\begin{equation}}
\newcommand{\eeq}{\end{equation}}
\newcommand{\bea}{\begin{eqnarray}}
\newcommand{\eea}{\end{eqnarray}}

\newcommand{\bean}{\begin{eqnarray*}}
\newcommand{\eean}{\end{eqnarray*}}

\newcommand{\Ref}[1]{(\ref{#1})}
\newcommand{\prd}{Phys. Rev. D\ }

\begin{document}

\begin{center}
{\Large\bf Cosmological evolution of a ghost scalar field}
\vskip12pt
Sergey V. Sushkov\footnote{sergey$\_$sushkov@mail.ru}\\
{\it Department of Mathematics, Kazan State Pedagogical
University, Mezhlauk 1 st., Kazan 420021, Russia}\\
\vskip6pt
Sung-Won Kim\footnote{sungwon@mm.ewha.ac.kr}\\
{\it Department of Science Education, Ewha Womans University,
Seoul 120-750, Korea}

\vskip12pt
\parbox{10cm}{\small We consider a scalar field with a
negative kinetic term minimally coupled to gravity. We obtain an
exact non-static spherically symmetric solution which describes a
wormhole in cosmological setting. The wormhole is shown to connect
two homogeneous spatially flat universes expanding with
acceleration. Depending on the wormhole's mass parameter $m$ the
acceleration can be constant (the de Sitter case) or infinitely
growing.}

\end{center}

\section{Basic equations}
Consider a real scalar field $\phi$ minimally coupled to general
relativity with the action given as follows
\beq\label{action}
S=\int d^4x\sqrt{-g}\left[\frac1{8\pi G}R -\varepsilon(\nabla\phi)^2-
2V(\phi)\right],
\eeq
where $g_{\mu\nu}$ is a metric, $g=\det(g_{\mu\nu})$, and $R$ is the
scalar curvature. We will take the potential $V(\phi)$ in the exponential
form:
\beq
V=V_0\exp(-k\phi).
\eeq
The exponential potential has been considered in numerous papers
devoted to cosmological models with scalar fields (see, for
instance,
\cite{Hal,Bar,BarGot,GotSai,BurBar,FeiIba,AguFeiIba,LucMat,LidMazSch,
MalWan,CopMazNun,Fin,Fonarev}). It arises as an effective
potential in some supergravity theories or in Kaluza-Klein
theories after dimensional reduction to an effective
four-dimensional theory \cite{Hal}. The exponential potential also
arises in higher-order gravity theories after a transformation to
the Einstein frame \cite{Bar,BarGot,GotSai,BurBar}. We will assume
that a parameter $\varepsilon$ in the action \Ref{action} can take
two values $\pm1$. The choice $\varepsilon=1$ corresponds to an
ordinary scalar field. The static spherically symmetric solution
of the theory \Ref{action} with $\varepsilon=1$ was obtained in
\cite{Buc,JanRobWin}. With a view of cosmological applications
this theory was studied in \cite{EllMad,HusMarNun,Fonarev}. The
choice $\varepsilon=-1$ gives a scalar field with a negative
kinetic term or so-called ghost scalar field. Scalar fields with
the opposite sign of their kinetic terms have been previously
considered in the literature. They appear in certain models of
inflation \cite{ArmDamMuk,ArmMukSte}, they have been proposed as
dark energy candidates \cite{Cal,SchWhi}, and they also appear in
certain unconventional supergravity theories which admit de Sitter
solutions \cite{Hul}. Moreover, ghost scalar fields have been
shown to allow wormhole solutions
\cite{Ell,Bro,Kod,Arm,HayKimLee}.

In this work we will find an exact cosmological solution with a
ghost scalar field in the theory \Ref{action}. Therefore
hereinafter we will assume $\varepsilon=-1$. In this case varying
the action \Ref{action} gives the Einstein equations
    \beq\label{Einstein}
    R_{\mu\nu}=8\pi
    G\left[-\nabla_\mu\phi\nabla_\nu\phi+g_{\mu\nu}V_0\exp(-k\phi)\right],
    \eeq
and the equation of motion of the scalar field
    \beq\label{eqmo}
    \nabla^\alpha\nabla_\alpha\phi=kV_0\exp(-k\phi).
    \eeq

\section{Static solution}
In the case $V_0=0$ (no potential term) the static solution to
Eqs. \Ref{Einstein}, \Ref{eqmo} was obtained by Bronnikov
\cite{Bro} and more recently by Armend\'ariz-Pic\'on \cite{Arm}.
Adopting the result of \cite{Arm} we can write down the solution
as follows
\beq\label{staticmetric}
ds^2=-e^{2u(r)}dt^2+e^{-2u(r)}\left[dr^2+(r^2+r_0^2)d\Omega^2\right],
\eeq
\beq\label{staticsf}
\phi(r)=(4\pi G\alpha^2)^{-1/2}u(r),
\eeq
where $d\Omega^2=d\theta^2+\sin^2\theta d\varphi^2$ is the metric of the
unit sphere, the radial coordinate $r$ varies from $-\infty$ to $\infty$,
and
\beq\label{u}
u(r)=\frac{m}{r_0}\arctan{\frac{r}{r_0}},
\eeq
\beq\label{alpha}
\alpha^2=\frac{m^2}{m^2+r_0^2},
\eeq
with $m$ and $r_0$ being two free parameters. Taking into account the
following asymptotical behavior:
\beq\label{asyu}
e^{2u(r)}=\exp\left(\pm\frac{\pi
m}{r_0}\right)\left[1-\frac{2m}{r}\right]+O(r^{-2})
\eeq
in the limit $r\to\pm\infty$, we may see that the spacetime with
the metric \Ref{staticmetric} possesses by two asymptotically flat
regions. These regions are connected by the throat whose radius
corresponds to the minimum of the radius of two-dimensional
sphere, $R^2(r)=e^{-2u(r)}(r^2+r_0^2)$. The minimum of $R(r)$ is
achieved at $r=m$ and equal to
\beq
R_{0}=\exp\left(-\frac{m}{r_0}\arctan\frac{m}{r_0}\right)(m^2+r_0^2)^{1/2}.
\eeq
Asymptotical masses, corresponding to $r\to\pm\infty$, are equal
to $m_{\pm}=\pm m\exp(\pm\pi m/2r_0)$. Note that the masses have
both different values and different signs. A behavior of the
scalar field is given by Eq. \Ref{staticsf}; it is seen that the
scalar field smoothly varies between two asymptotical values
$\phi_\pm=\pm(\pi/16G)^{1/2}[1+(m/r_0)^2]^{1/2}$.  Thus, we may
summarize that the metric \Ref{staticmetric} describes a static
spherically symmetric wormhole with the throat's radius $R_0$ and
two different asymptotical masses $m_\pm$. When $m=0$, the static
solution \Ref{staticmetric}, \Ref{staticsf} reduces to
\beq\label{MTwh}
ds^2=-dt^2+dr^2+(r^2+r_0^2)d\Omega^2,
\eeq
\beq
\phi(r)=\frac1{\sqrt{4\pi G}}\arctan{\frac{r}{r_0}}.
\eeq
It is worth noting that the metric \Ref{MTwh} was proposed {\em a priori}
by Morris and Thorne in the pioneering work \cite{MorTho} as a simple
example of the wormhole spacetime metric.

\section{Non-static solution}
Let us now consider a non-static spherically symmetric solution of the
equations \Ref{eqmo}, \Ref{Einstein} in the case $V_0\not=0$. Such the
solution is given by the following statement:

\vskip6pt\noindent {\bf Statement.}\ A time-dependent spherically
symmetric solution of \Ref{eqmo}, \Ref{Einstein} exists if and
only if $V_0>0$ and, in this case, the general solution has the
following form: \beq\label{nonstaticmetric} ds^2=-\exp(-2\alpha^2
aT+2u)\,dT^2 +\exp(2aT-2u)\left[dr^2+(r^2+r_0^2)d\Omega^2\right],
\eeq \beq\label{nonstaticsf} \phi(T,r)=(4\pi
G\alpha^2)^{-1/2}\left[u-\alpha^2 aT\right], \eeq where $u(r)$ and
$\alpha$ are given by Eqs. \Ref{u}, \Ref{alpha}, and the
parameters $a$ and $\alpha$ are related to the parameters of the
potential as follows: \beq\label{param} k=4\alpha(\pi
G)^{1/2},\quad V_0=\frac{a^2(3+\alpha^2)}{8\pi G}. \eeq

\vskip6pt\noindent {\em Proof.} Let $\overline{g}_{\mu\nu}$ and
$\overline{\phi}$ be the `old' static solutions \Ref{staticmetric},
\Ref{staticsf}. Now consider the conformal transformation of the metric
\beq\label{newg}
g_{\mu\nu}=\exp(2\mu(t))\overline{g}_{\mu\nu},
\eeq
and suppose that at the same time the scalar field transforms as follows
\beq\label{newphi}
\phi=\overline{\phi}-(4\pi G)^{-1/2}\alpha\mu(t),
\eeq
where $\mu(t)$ is a new indefinite function of $t$. Using the
corresponding transformational properties of the Ricci tensor:
\bea
R_{00}&=&\overline{R}_{00}-3\ddot{\mu},\nonumber\\
R_{0i}&=&\overline{R}_{0i}+\dot{\mu}\partial_i\ln(g_{00}),\\
R_{ij}&=&\overline{R}_{ij}-(\ddot{\mu}+2\dot{\mu}^2)g_{ij}g^{00},\nonumber
\eea
and taking into account that $\overline{g}_{\mu\nu}$ and $\overline{\phi}$
satisfy the Einstein equations
\beq
\overline{R}_{\mu\nu}=-8\pi
G\overline{\phi}_{,\mu}\overline{\phi}_{,\nu},
\eeq
it is easy to check that the metric tensor \Ref{newg} and the
scalar field \Ref{newphi} satisfy the equation \Ref{Einstein}
provided $k=4\alpha(\pi G)^{1/2}$ and the function $\mu(t)$ obeys
the following two equations:
\bea
&&\ddot{\mu}=(1+\alpha^2)\dot{\mu}^2, \\
&&(3+\alpha^2)\dot{\mu}^2=8\pi
GV_0\exp\left[2(1+\alpha^2)\mu\right].
\eea
The solution of these equations is
\beq\label{solmu}
\mu(t)=-(1+\alpha^2)^{-1}\ln|(1+\alpha^2)at|,
\eeq
where $a$ is a free parameter. Substituting $\mu(t)$ given by Eq.
\Ref{solmu} into \Ref{newg}, \Ref{newphi}, we get
\beq\label{confmet}
ds^2=|(1+\alpha^2)at|^{-2/(1+\alpha^2)}
\left\{-e^{2u}dt^2+e^{-2u}\left[dr^2+(r^2+r_0^2)d\Omega^2\right]\right\},
\eeq
\beq
\phi(t,r)=(4\pi
G\alpha^2)^{-1/2}\left[u+\alpha^2(1+\alpha^2)^{-1}\ln|(1+\alpha^2)at|\right].
\eeq
And, at last, redefining the time coordinate,
\beq
\frac1{(1+\alpha^2)at}=\pm\exp((1+\alpha^2)aT),
\eeq
we arrive at \Ref{nonstaticmetric}, \Ref{nonstaticsf}.

To complete the proof we consider the scalar field equation \Ref{eqmo}.
Substituting the expression \Ref{newphi} into \Ref{eqmo} and taking into
account that $\overline{\phi}$ satisfies the equation
$\overline{\nabla}^{\alpha}\overline{\nabla}_{\alpha}\overline{\phi}=0$ we
find
\beq\label{eqmomu}
\ddot{\mu}+2\dot{\mu}^2=\frac{\sqrt{4\pi G}}{\alpha}kV_0
\exp\left[\left(2+\frac{k\alpha}{\sqrt{4\pi G}}\right)\mu\right].
\eeq
As is easy to check straightforwardly, this equation is valid for
$\mu(t)$, $k$, and $V_0$ given by the relations \Ref{solmu} and
\Ref{param}.\footnote{Notice also that the equation \Ref{eqmomu}
coincides, if $k=4\alpha(\pi G)^{1/2}$, with the $(1,1)$-component of
Einstein's equations.}\hfill$\Box$

\section{Wormhole in cosmological setting}
In this section we shall analyze the solution
\Ref{nonstaticmetric}, \Ref{nonstaticsf} found in the preceding
section. First, it is a solution of the Einstein-minimal coupling
scalar field equations with the potential $V(\phi)=W(T,r)$, where
\beq\label{W}
W(T,r)=\frac{a^2(3+\alpha^2)}{8\pi G}\exp(-2u+2\alpha^2 aT).
\eeq
The solution depends on three parameters $m$, $r_0$ and $a$. When $a=0$,
we get the static solution \Ref{staticmetric}, \Ref{staticsf} obtained in
\cite{Bro,Arm}. Depending on a value of $m$ we have two qualitatively
different cases:

\vskip6pt\noindent{\bf A.} The case $m=0$ ($\alpha=0$). The
solution \Ref{nonstaticmetric}, \Ref{nonstaticsf} now takes the
simple form:
\beq\label{metricm0}
ds^2=-dT^2+e^{2aT}[dr^2+(r^2+r_0^2)\,d\Omega^2],
\eeq
\beq
\phi(r)=\frac{1}{\sqrt{4\pi G}}\arctan\frac{r}{r_0}.
\eeq
Note that in this case the scalar field $\phi$ does not depend on
the time coordinate $T$, though the metric \Ref{metricm0} is
non-static. The potential \Ref{W} becomes to be constant:
\beq
W(r,T)\equiv\frac{3a^2}{8\pi G},
\eeq
and corresponds, in fact, to the positive cosmological constant
$\Lambda=3a^2$ in the action \Ref{action}. It is easy to see that
at each moment of time the metric \Ref{metricm0} coincides
asymptotically (i.e. in the limit $r\to\pm\infty$) with the de
Sitter one, and an intermediate region represents a throat
connecting these asymptotically de Sitter regions. Thus, the
spacetime \Ref{metricm0} is a wormhole joining two de Sitter
universes. The instant radius of the throat is equal to the
minimal radius of two-dimensional sphere, $R_0=e^{aT}r_0$; we see
that it grows exponentially with time. Let us calculate now the
scalar curvature:
\beq
R=12a^2-\frac{2r_0^2e^{-2aT}}{(r^2+r_0^2)^2}.
\eeq
In the limit $r\to\pm\infty$ as well as in the limit $T\to\infty$
the scalar curvature has the De-Sitter value $R_{DS}=12a^2$, while
at $T=-\infty$ the scalar curvature is singular. This singularity
has a clear geometrical interpretation. Namely, at each moment of
time the throat is represented as the 2D sphere of minimal radius.
In the limit $T\to-\infty$ the radius of sphere $R_0=e^{aT}r_0$
tends to zero, the curvature of sphere goes to infinity, and the
corresponding spacetime scalar curvature $R$ becomes to be
singular.

It is worth also noting that a metric of the kind of
\Ref{metricm0} was first introduced {\em a priori}\ by Roman
\cite{Rom:93}, who explored the possibility that inflation might
provide a mechanism for the enlargement of submicroscopic, i.e.,
Planck scale wormholes to macroscopic size.

\vskip6pt\noindent{\bf B.} The case $m\not=0$ ($\alpha\not=0$). In
this case the solution is described by the general formulas
\Ref{nonstaticmetric}, \Ref{nonstaticsf}, and the potential
$V(\phi)=W(r,T)$ is given by \Ref{W}. The scalar curvature
calculated in the metric \Ref{nonstaticmetric} reads
\beq
R=6a^2(2+\alpha^2)e^{-2u+2\alpha^2 aT}
-\frac{2(m^2+r_0^2)e^{2u-2aT}}{(r^2+r_0^2)^2}.
\eeq
As in the case $m=0$ the scalar curvature is singular at
$T=-\infty$, and in addition it is now blowing up at $T\to\infty$.
To characterize the last singularity it will be convenient to
introduce the proper time $\tau$ as
\beq
-\alpha^2a\tau=\exp(-\alpha^2aT),
\eeq
so that $\tau$ runs from $-\infty$ to $0_-$ while $T$ varies from
$-\infty$ to $+\infty$. Now we may rewrite the metric
\Ref{nonstaticmetric} as follows
\beq
ds^2=-e^{2u}d\tau^2+|\alpha^2
a\tau|^{-2/\alpha^2}e^{-2u}\left[dr^2+(r^2+r_0^2)d\Omega^2\right].
\eeq
In the limit $r\to\pm\infty$ the last metric describes an
homogeneous spatially flat universe:\footnote{In order to obtain
Eq. \Ref{asymet} we must take into account the asymptotical
formula \Ref{asyu} and make the rescaling
$\tilde\tau=\tau\exp(\pm\pi m/2r_0)$ and $\tilde
r=r[\alpha^2a\exp(\mp\pi m/2r_0)]^{-1/\alpha^2}$.}
\beq\label{asymet}
ds^2=-d\tilde\tau^2+|\tilde\tau|^{-2/\alpha^2}\left[d{\tilde
r}^2+{\tilde r}^2 d\Omega^2\right],
\eeq
with the scale factor $a(\tilde\tau)=|\tilde\tau|^{-1/\alpha^2}$
and the scalar curvature
$$R =\frac{6(2+\alpha^2)}{\alpha^4\tilde\tau^2}.$$
The corresponding Hubble parameter $\dot{a}/a$ is equal to
$|\alpha^2\tilde\tau|^{-1}$, and the acceleration parameter
$\ddot{a}a/\dot{a}$ is $(1+\alpha^2)(\alpha^4\tilde\tau^2)^{-1}$.
It is seen that the universe is expanding with an acceleration
into a ``final'' singularity at $\tilde\tau=0_-$, and the Hubble
and acceleration parameters are infinitely growing in the course
of expansion.

Thus, in the case $m\not=0$, the metric \Ref{nonstaticmetric}
represents a wormhole connecting two homogeneous spatially flat
universes expanding with infinitely growing acceleration.

\section{Summary}
In this paper we have obtained the exact non-static spherically
symmetric solution \Ref{nonstaticmetric}, \Ref{nonstaticsf} in the
theory of gravity with the ghost scalar field possessing the
exponential potential. The spacetime described by the metric
\Ref{nonstaticmetric} represents two asymptotically homogeneous
spatially flat universes connected by a throat. In the other
words, one may interpret such the spacetime as a wormhole in
cosmological setting. It is important to notice that both the
universes and the throat of the wormhole are simultaneously
expanding with acceleration. The character of acceleration
qualitatively depends on the wormhole's mass parameter $m$. In
case $m=0$ the acceleration is constant, so that the corresponding
spacetime configuration, given by the metric \Ref{metricm0},
represents two de Sitter universes joining by the throat. Note
that Roman \cite{Rom:93} has considered such the spacetime as an
example of inflating wormholes. In case $m\not=0$ the acceleration
turns out to be infinitely growing, so that the metric
\Ref{nonstaticmetric} describes now the inflating wormhole
connecting two homogeneous spatially flat universes expanding with
infinitely growing acceleration into the final singularity.

\section*{Acknowledgments}
S.S. acknowledge kind hospitality of the Ewha Womans University.
S.S. was also supported in part by the Russian Foundation for
Basic Research grant No 02-02-17177. S.-W.K. was supported in part
by grant No. R01-2000-00015 from the Korea Science \& Engineering
Foundation.


\begin{thebibliography}{99}
\bibitem{Hal}
J. J. Halliwell, Phys. Lett. {\bf 175B}, 341 (1987).
\bibitem{Bar}
J. D. Barrow, Nucl. Phys. {\bf B 296}, 697 (1988).
\bibitem{BarGot}
J. D. Barrow, S. Gotsakis, Phys. Lett. {\bf 214B}, 515 (1988);
Phys. Lett. {\bf 258B}, 299 (1988).
\bibitem{GotSai}
S. Gotsakis, P. J. Saich, Class. Quantum Grav. {\bf 11}, 383
(1994).
\bibitem{BurBar}
A. B. Burd, J. D. Barrow, Nucl. Phys. {\bf B 308}, 929 (1988).
\bibitem{LucMat} F. Lucchin, S. Mataresse, \prd {\bf 32}, 1316
(1985).
\bibitem{FeiIba}
A. Feinstein, J. Ib\'a\~nez, Class. Quantum Grav. {\bf 10}, 93,
L227 (1993).
\bibitem{AguFeiIba}
J. M. Aguirregabiria, A. Feinstein, J. Ib\'a\~nez, \prd {\bf 48},
4662, 4669 (1993).
\bibitem{LidMazSch}
A. R. Liddle, A. Mazumdar, F. E. Schunck, \prd {\bf58}, 061301
(1998).
\bibitem{MalWan}
K. Malik, D. Wands, \prd {\bf59}, 123501 (1999).
\bibitem{CopMazNun}
E. J. Copeland, A. Mazumdar, N. J. Nunes, \prd {\bf60}, 083506
(1999).
\bibitem{Fin}
F. Finelli, Phys. Lett. {\bf 545B}, 1 (2002);
arXiv:hep-th/0206112.
\bibitem{Buc}
H. Buchdal, Phys. Rev. {\bf 115}, 1325 (1959).
\bibitem{JanRobWin}
A. I. Janis, D. C. Robinson, J. Winicour, Phys. Rev. {\bf186},
1729 (1969).
\bibitem{EllMad}
G. F. R. Ellis, M. S. Madsen, Class. Quantum Grav. {\bf 8}, 667
(1991).
\bibitem{HusMarNun}
V. Husain, E. A. Martinez, D. N\'u\~nez, \prd {\bf 50}, 3783
(1994).
\bibitem{Fonarev}
O. A. Fonarev, Class. Quantum Grav. {\bf 12}, 1739 (1995).
\bibitem{ArmDamMuk}
C. Armend\'ariz-Pic\'on, T. Damour, V. Mukhanov, Phys. Lett. B
{\bf 458}, 209 (1999).
\bibitem{ArmMukSte}
C. Armend\'ariz-Pic\'on, V. Mukhanov, P. Steinhardt, \prd {\bf
63}, 103510 (2001).
\bibitem{Cal}
R. Caldwell, Phys. Lett. B {\bf 545}, 23 (2002).
\bibitem{SchWhi}
A. Schulz, M. White, \prd {\bf 64}, 043514 (2001).
\bibitem{Hul}
C. Hull, J. High Energy Phys. {\bf 11}, 012 (2001).
\bibitem{Ell}
H. Ellis, J. Math. Phys. {\bf 14}, 104 (1973).
\bibitem{Bro}
K. A. Bronnikov, Acta Physica Polonica {\bf B 4}, 251 (1973).
\bibitem{Kod}
T. Kodama, \prd {\bf 18}, 3529 (1978).
\bibitem{Arm} 
C. Armend\'ariz-Pic\'on, \prd {\bf 65}, 104010 (2002).
\bibitem{HayKimLee}
S. A. Hayward, S.-W. Kim, and H. Lee, \prd {\bf 65}, 064003
(2002).
\bibitem{MorTho}
M. S. Morris, K. S. Thorne, Am. J. Phys. {\bf 56}, 395 (1988).
\bibitem{Rom:93}
T. A. Roman, \prd {\bf 47}, 1370 (1993).

\end{thebibliography}
\end{document}